\documentclass[a4paper,12pt,openright]{book}
\usepackage{amsmath, amsthm, amssymb, bm, braket}
\usepackage[dvips]{graphicx}
\usepackage{cite, caption, wrapfig, fancybox, fancyhdr, setspace}
\usepackage{listings}%---for displaying the codes.
\usepackage{color}
\usepackage{geometry}
  \renewcommand{\chaptermark}[1]{\markboth{Chapter~\thechapter\; \; #1}{}}
\begin{document}
\renewcommand{\figurename}{Figure}
\renewcommand{\tablename}{Table}
\renewcommand{\thetable}{\arabic{table}}
\newcommand{\bi}[1]{\ensuremath{ \boldsymbol{#1} }}
\newcommand{\oprt}[1]{\ensuremath{ \hat{\mathcal{#1}} }}
\newcommand{\abs}[1]{\ensuremath{ \left| #1 \right| }}
\newcommand{\cgc}[2]{\ensuremath{ \mathcal{C}^{#1}_{#2} }}
\newcommand{\cmd}[1]{\texttt{\symbol{"5C}#1}}
\def \Schr{Schr\"odinger }
\def \CG{Clebsch-Gordan }
\def \beq{\begin{equation}}
\def \eeq{\end{equation}}
\def \beqa{\begin{eqnarray}}
\def \eeqa{\end{eqnarray}}
\def \bis{\bi{s}}
\def \bir{\bi{r}}
\def \ubir{\bar{\bi{r}}}
\def \bip{\bi{p}}
\def \ubip{\bar{\bi{r}}}
%%%%%%%%%%%%%%%%%%%%%%%%%%%%%%%%%%%%%%%%%%%%%%%%%%%%%%%%%%%%%%%%%%%%

%\renewcommand{\include}[1]{} %annul ``include'' command.
%\renewcommand{\documentclass}[2][]{} %annul ``\documentclass'' command.
\title{Instruction of my personal computing library}
\author{Tomohiro Oishi}
\date{11th April 2023}
\maketitle

{\bf Abstract}

This document is prepared to introduce and explain how to use the computing library composed by T. Oishi.
The main purpose of library is to perform numerical calculations in the nuclear physics.
Its final version is expected to be published for educational and commercial purposes.
Before the official publication, under the agreement with publishers,
I make the current, preliminary version open for public.
For full usage, two applications, GFORTRAN and GNUPLOT, are necessary.
Feedbacks and comments on products will be appreciated.
The source codes etc. are available in the GitHub repository \cite{2023Oishi_GH}.

\vspace{20truemm}

{\bf Acknowledgment}

For composing this library,
the computer with Ubuntu LINUX and several free applications,
including GFORTRAN, GNUPLOT, EMACS, etc., are utilized.
I am thankful for all, who had contributions to these products.
I would like to give my appreciation to the following people,
who supervised and/or encouraged this work:
John Smith,
Gonpei Nanashino.

\tableofcontents

\chapter{Library-01: TOSPEM} \label{Ch_TOSPEM}
The code TOSPEM written in fortran-90 solves,
for the spherical nucleus of $(Z,N)$ protons and neutrons,
(i) the Schr\"{o}dinger equation for the single-nucleon states within the Woods-Saxon potential,
(ii-a) the electric or magnetic transition strength, $B_{EJ}$ or $B_{MJ}$, between
the arbitrary set of initial and final states of the nucleus of interest, and
(ii-b) Weisskopf estimate for comparison with results in (ii-a) \cite{1951Weiss}.
The similar code was utilized in my past works \cite{2011Oishi,2019OP}.

This note and the code TOSPEM.f90 are based on the CGS-Gauss system of units.
There exists practical difference between the CGS-Gauss and SI (MKS-Ampere) systems
but only in the electro-magnetic quantities.
See TABLE \ref{table:units} for details.

%%%%%%%%%%%%%%%%%%%%%%%%%%%%%%%%%%%%%%%%%%%%%%%%%%%%%%%%%%%%%%%%%%%%%%%%%%%%%%%%%%%%%%%%%
\section{structure of code}
The code TOSPEM.f90 includes the following parts.
\begin{itemize}
\item ``HONTAI'' as the main part.
\item ``mdl\_XXX\_*'' as the modules, including parameters, functions, and subroutines.
\item ``SPEM'' for (i) solving the single-particle Sch\"{o}dinger equation, and (ii) computing the $B_{EJ}$ or $B_{MJ}$ based on the formulas by Suhonen, Ring, and Schuck \cite{2007Suhonen,80Ring}.
\item ``Weisskopf'' for Weisskopf's estimate of $B_{EJ}$ or $B_{MJ}$.
\item There is the external file ``PARAM.inp'' for several input parameters. The quantum labels, $\left\{nlj \right\}$ for the initial and final states, and gyro-magnetic factors, $g^{(p)}_{s/l}$ and $g^{(n)}_{s/l}$ for $M\lambda$ modes, are determined there.
\end{itemize}
Here the several TIPs for using this code.
\begin{itemize}
\item For compiling and executing the code, ``A\_comprunshow.sh'' (shell-script file for BASH) is prepared.
\item Cutoff parameters $E_{cut}$, $l_{max}$, $r_{\rm max}$, and $dr$ are fixed in the module, mdl\_001\_setting. If you change them, compile again the code.
\item For determining the system of interest, fix these parameters: (i) zc and nc in mdl\_001\_setting; (ii) parameters r00, W\_0, etc. in mdl\_002\_potentials. Also, modify the input parameters in ``PARAM.inp'' as necessary.
\item In the module, mdl\_002\_potentials, single-particle potentials are determined as functions, v\_cp and v\_cn. Those are saved in output files, ``Yed\_XXXX\_SP\_Potentials.dat'', for several $(l,j)$ channels.
\end{itemize}

%%%%%%%%%%%%%%%%%%%%%%%%%%%%%%%%%%%%%%%%%%%%%%%%%%%%%%%%%%%%%%%%%%%%%%%%%%%%%%%%%%%%%%%%%
\section{single-nucleon states}
In the first part of this code \textsc{TOSPEM}, the spherical Schr\"{o}dinger
equation is solved within the Woods-Saxon potential, which is determined in
the fortran-90 module, \textsc{mdl\_002\_potentials}.
Namely, the single-nucleon wave function satisfies
\beq
\left[ -\frac{\hbar^2}{2\mu} \nabla_{\bir}^2 +V(r) \right] \psi(\bir) = E \psi(\bir).
\eeq
The spherical solution generally reads
\beq
\psi_{nljm}(\bir) = u_{nlj}(r) \mathcal{Y}_{ljm}(\ubir) \equiv \frac{A(r)}{r} \mathcal{Y}_{ljm}(\ubir),
\eeq
where $\left\{ n,l,j,m=-j \sim j \right\}$ indicate the radial node, orbital angular momentum,
coupled angular momentum, and magnetic quantum number, respectively.
Since spherical, what the code needs to compute is only the radial part $u_{nlj}(r)$.
For example, the $0s_{1/2}$ state is solved as $u_{00\frac{1}{2}}(r)$.
For numerical solution, the energy cutoff $ecut$ and the radial box $r_{\rm max}$ are
fixed in the module \textsc{mdl\_001\_setting}.
Note also that each state is normalized in the code:
$\int d \bir \abs{\psi_{nljm}(\bir)}^2 =1$.

By using $v(r) \equiv 2\mu V(r)/\hbar^2$ and $\epsilon \equiv 2\mu E/\hbar^2$,
the original equation changes as
\beqa
\left[ -\left( r^{-1}\frac{d^2}{dr^2}r -\frac{l(l+1)}{r^2} \right) +v(r)-\epsilon \right] \frac{A(r)}{r} &=& 0 \nonumber \\
\left[ \frac{d^2}{dr^2}  -\frac{l(l+1)}{r^2} -v(r) +\epsilon \right] A(r) &=& 0 \nonumber  \\
\frac{d^2}{dr^2} A(r) = G_E(r) A(r), \label{eq:VHSUYTE} &&
\eeqa
where
\beq
G_E(r) = v(r) -\epsilon +\frac{l(l+1)}{r^2} =\frac{2\mu V(r)}{\hbar^2} -\frac{2\mu E}{\hbar^2} +\frac{l(l+1)}{r^2}.
\eeq
Thus, when $V(r\rightarrow 0)$ is not divergent, the asymptotic solutions read
\beq
A(r\rightarrow 0) \cong  r^{l+1},~~~A'(r\rightarrow 0) \cong (l+1) r^{l+1}. \nonumber
\eeq
Also, when $V(r\rightarrow \infty)=0$,
\beq
A(r\rightarrow \infty) \cong re^{-kr},~~~A'(r\rightarrow \infty) \cong (1-kr)e^{-kr}, \nonumber
\eeq
where $k=\sqrt{-2\mu E} /\hbar$ for the bound state with $E<0$.
These asymptotic forms help us to infer whether the numerical routine
is correctly installed: if there is a problem, the output wave functions
do not behave like these asymptotic forms.

For the radial part, $u_{nlj}(r) = \frac{A_{nlj}(r)}{r}$, the function $A_{nlj}(r)$
is computed with ``Numerov method'' \cite{93Hairer}.
This method generally solves an equation in the form of
\beq
\left[ \frac{d^2}{dr^2} + w(r) \right] A(r) =0.
\eeq
Then, if one can determine the first two points,
$A_0 \equiv A(r_0)$ and $A_1 \equiv A(r_1)$ from e.g. the asymptotic forms,
the remaining points are computed as
\beq
A_{n+1} \cong \frac{(2-5a^2 w_n/6)A_n  -(1+a^2 w_{n-1})A_{n-1}}{1+a^2 w_{n+1}/12},
\eeq
where $a$ is the radial mesh.
For more details of Numerov method, see Appendix \ref{Ap_Numerov}.

In the code, for finding the correct eigen energy of the bound state,
the node-counting procedure is used.
Namely, the eigen energy $E_{nlj}$ is determined as the maximum energy but keeping
the node number of $u_{nlj}(r)$ as $n$.
There are two subroutines of Numerov methods starting from
$r=0$ fm and $r=r_{\rm max}$.
Matching of these forward and backward solutions is necessary for bound states.
The continuum-energy levels $E_{nlj}>0$, on the other hand, are discretized within the
box-boundary condition, namely, to satisfy that $u_{nlj}(r_{\rm max})=0$.

\section{sample calculation of single-nucleon states} \label{sec:WERTYEFWQ}
Here I introduce the results of sample calculations for the $^{40}$Ca nucleus.
Note that the radial-wave functions, $A_{nlj}(r) \equiv r \cdot u_{nlj}(r)$, are stored as the array ``psi''.
Those are normalized as $\int dr r^2 \abs{u_{nlj}(r)}^2 =\int dr \abs{A_{nlj}(r)}^2 =1$.

\subsection{neutron states}
The single-particle Schr\"{o}dinger equation for neutrons is given as 
\beq
\left[ -\frac{\hbar^2}{2\mu} \nabla_{\bir}^2 +V(r) \right] \psi(\bir) = E \psi(\bir),
~~~~
\psi_{nljm}(\bir) = u_{nlj}(r) \mathcal{Y}_{ljm}(\ubir),
\eeq
where $\mu = m_n m_c/(m_n +m_c)$, $m_c = 20m_n + 20m_p -40B/c^2$,
$m_n c^2 =939.5654133$ MeV, $m_p c^2 =938.2720813$ MeV, and
$B=8.551305$ MeV, which is the binding energy per nucleon of $^{40}$Ca \cite{NNDC_Chart}.
The Woods-Saxon potential reads
\beqa
&& V(r) = V_{WS}(r) = V_0 f(r) + U_{ls} (\bi{l} \cdot \bi{s}) \frac{1}{r} \frac{df(r)}{dr}, \nonumber  \\
&& f(r) = \frac{1}{1 + e^{(r-R_0)/a_0}}, \label{eq:ch1_WS}
\eeqa
where $R_0=r_0\cdot 40^{1/3}$, and $f(r)$ is the standard Fermi profile.
In this library, I fix the parameters as $V_0=-55.57$ MeV,
$U_{ls}=11.28$ MeV$\cdot$fm$^2$, $r_0 =1.25$ fm, and $a_0 =0.65$ fm.
Notice that the spin-orbit (LS) term is included.
In addition, I employ the cutoff parameters, $r_{max}=30$ fm, $dr=0.1$ fm, $l_{max}=5$, and $E_{max}=18$ MeV.
The results are obtained as follows.
\begin{lstlisting}[basicstyle=\ttfamily\footnotesize, frame=single]
 (neutron-core) s.p. states
    # Node    L    J*2      E(MeV)
    1    0    0    1    -42.6596966
    2    0    1    3    -32.0630613
    3    0    1    1    -31.2326299
    4    0    2    5    -20.4627846
    5    0    2    3    -18.7572561
    6    1    0    1    -17.7700038
    7    0    3    7     -8.3626685
    8    1    1    3     -6.2390782
    9    0    3    5     -5.7572462
   10    1    1    1     -5.4529772
   11    2    0    1      0.1683648
   12    2    1    3      0.5043552
   13    2    1    1      0.5071959
   14    1    2    5      0.7622601
   15    1    2    3      0.7713305
   16    3    0    1      0.8443721
   17    1    3    7      1.1522559
   18    1    3    5      1.1525952
   19    3    1    3      1.5695544
   20    0    4    9      1.5799132
   89    8    2    3     17.8490518
 number of s.p.basis=          89
\end{lstlisting}

\subsection{proton states}
For proton states, the single-particle Schr\"{o}dinger equation should include
the repulsive Coulomb potential.
In this library, the potential of uniformly-charged spherical core is employed.
Namely,
\beq
V(r) = V_{WS}(r) +V_{C}(r),~~~~V_{C}(r) = \left\{ \begin{array}{ll} \frac{Ze^2}{r}  &(r \ge R_0), \\  \frac{Ze^2}{2R_0}\left[3-\left( \frac{r}{R_0} \right)^2  \right] &(r < R_0), \end{array} \right.
\eeq
with $R_0 = r_0 \cdot 40^{1/3}$.
Also the relative mass should be slightly modified as $\mu = m_p m_c/(m_p +m_c)$.
Then the results are obtained as follows.
Notice that each proton's energy is higher than the neutron's energy, because of the repulsive Coulomb potential.
Several levels become unbound in the proton side.
\begin{lstlisting}[basicstyle=\ttfamily\footnotesize, frame=single]
 (proton-core) s.p. states
    # Node    L    J*2      E(MeV)
    1    0    0    1    -33.6446148
    2    0    1    3    -23.5993700
    3    0    1    1    -22.7430551
    4    0    2    5    -12.5323720
    5    0    2    3    -10.7995723
    6    1    0    1     -9.8024169
    7    0    3    7     -1.0297129
    8    1    1    3      0.6845982
    9    1    1    1      1.3588870
   10    0    3    5      1.5319337
   11    2    0    1      1.9225865
   12    2    1    3      2.0435195
   13    2    1    1      2.0490777
   14    1    2    5      2.2519461
   15    1    2    3      2.2520630
   16    1    3    7      2.5540474
   17    1    3    5      2.5543557
   18    0    4    9      2.9317977
   19    0    4    7      2.9318078
   20    3    0    1      3.2036884
   80    5    5   11     17.7374567
 number of s.p.basis=          80
\end{lstlisting}

\section{eclectic/magnetic transitions}
Electro-magnetic multi-pole transition of single proton or neutron inside
the atomic nucleus is described by the operator,
\beq
\oprt{Q} = \oprt{Q}(X\lambda \mu, \bir_{p/n}),
~~~{\rm for}~~~
\Braket{f\mid \oprt{Q} \mid i}=\int d\bir \psi^*_f (\bir) \oprt{Q}(X\lambda \mu, \bir) \psi_i(\bir),
\eeq
where $X=E~(M)$ for the electric (magnetic) mode. 
Its formalism is given as Eqs. (B.23) and (B.24) in the textbook \cite{80Ring}. 
Those are, 
\beqa
 \oprt{Q}(E\lambda \mu, \bir) &=& e_{\rm eff} r^{\lambda} Y_{\lambda \mu}(\ubir), \nonumber \\
 \oprt{Q}(M\lambda \mu, \bir) &=& \mu_{\rm N} \left( \vec{\nabla} r^{\lambda} Y_{\lambda \mu}(\ubir) \right) \cdot \left( \frac{2g_l}{\lambda+1}\bi{\hat{l}} + g_s\bi{\hat{s}} \right), \nonumber
\eeqa
where $e_{\rm eff}$, $\mu_{\rm N}$ (nuclear magneton), 
$g_l$, and $g_s$ are the well-known effective parameters \cite{2007Suhonen,80Ring}.
Usually, for the proton (neutron),
$e_{\rm eff}={\rm e}~(0)$, $g_l=1~(0)$, and  $g_s=5.586~(-3.826)$.
Also, the nuclear magneton is given as $\mu_{\rm N}= {\rm e} \hbar/2m_p c \cong 0.10515 \left[ {\rm e} \cdot{\rm fm} \right]$.

\begin{table*}[tb] \begin{center}
\caption{
Notations in the SI (MKS-Ampere) and CGS-Gauss systems of units especially for nuclear-physical quantities \cite{2007Suhonen,80Ring}.
Note also that $m_p\cong 938.272$~MeV$/c^2$ (proton mass), and fm$^2=10$~mb$=10^{-2}$~barn for the typical order of cross sections.
} \label{table:units}
 \catcode`? = \active \def?{\phantom{0}} %define `?' as ' '(one-blank).
 \begingroup \renewcommand{\arraystretch}{1.5}
 \begin{tabular*}{\hsize} { @{\extracolsep{\fill}} llll } \hline \hline
 ~Quantity            &SI~[unit]                     &CGS-Gauss~[unit]   &Note     ~\\ \hline
 ~Elementary charge  &$\equiv e$   &$\equiv {\rm e}$  &conversion:~\\
 ~                   &$\cong 1.602 \times 10^{-19}$[C]   & &$e^2 \longrightarrow 4\pi \epsilon_0 {\rm e}^2$~\\
 ~&&& ~\\
 ~Coulomb potential  &$V=\frac{1}{4\pi \epsilon_0}\frac{e^2}{r}$~[MeV]  &$=\frac{{\rm e}^2}{r}$~[MeV]  &between 2p~\\
 ~Fine-structure constant &$\alpha=\frac{e^2}{4\pi \epsilon_0 \hbar c}$  &$=\frac{{\rm e}^2}{\hbar c}$  &$\cong \frac{1}{137.036}$~\\
 ~Nuclear magneton &$\mu_{\rm N}=\frac{e \hbar}{2m_p}$  &$\mu_{\rm N}=\frac{{\rm e} \hbar}{2m_p c}$  &~\\
 ~    &$\cong 0.10515$~[$ce \cdot$fm]  &$\cong 0.10515$ [e$\cdot$fm]  &~\\
 ~&&& ~\\
 ~$B_{EJ}(E)$      &in $\left[e^2 {\rm fm}^{2J}  \right]$  &in $\left[{\rm e^2 fm}^{2J} \right]$ &~\\
 ~$B_{MJ}(E)$      &in $\left[ \frac{\mu^2_{\rm N}}{c^2} {\rm fm}^{2J-2}  \right]$  &in [$\mu^2_{\rm N} {\rm fm}^{2J-2}$] &~\\
 ~             &$\cong 1.106\times 10^{-2}\left[ e^2 {\rm fm}^{2J} \right]$  &$\cong 1.106\times 10^{-2}\left[ {\rm e}^2 {\rm fm}^{2J} \right]$  &~\\
 ~&&& ~\\
 ~$T_{EJ/MJ}(E)$
    &$=\frac{2}{\epsilon_0 \hbar} f(J) \left( \frac{E}{\hbar c} \right)^{2J+1}$ &$=\frac{8\pi}{\hbar} f(J) \left( \frac{E}{\hbar c} \right)^{2J+1}$   &$f(J)\equiv \frac{J+1}{J}$~\\
    &$\times B_{EJ/MJ}(E)$ &$\times B_{EJ/MJ}(E)$   &$\times \left( \frac{1}{(2J+1)!!} \right)^2$~\\
 \hline \hline
 \end{tabular*}
 \endgroup
 \catcode`? = 12 %initialize `?'.
\end{center} \end{table*}

Transition probability per time due to the electric/magnetic
transitions is formulated as Eq. (B.72) in Ref. \cite{80Ring}:
\beqa
 && T(X \lambda \mu;i\rightarrow f)
 = \frac{8\pi}{\hbar} f(\lambda) \left( \frac{E_{fi}}{\hbar c} \right)^{2\lambda+1} \times B(X \lambda \mu;i\rightarrow f)~~~[s^{-1}],~{\rm with}  \label{eq:TB} \\
 && f(\lambda) \equiv  \frac{\lambda+1}{\lambda} \frac{1}{[(2\lambda+1)!!]^2}, \nonumber
\eeqa
where $E_{fi}=E_f-E_i$ \footnote{Within the MKS-Ampere system of units, the first factor 
of Eq.(\ref{eq:TB}) should be replaced into $\frac{2}{\epsilon_0 \hbar}$.
Note also that the definition of $\mu_{\rm N}$ should be different from that in the CGS-Gauss system.}.
Here $B(i\rightarrow f)$ is the total transition strength, which is represented as
\beq
B(X \lambda \mu;i\rightarrow f) = \frac{1}{2I_i+1} \sum_{\mu M_i M_f} \abs{\Braket{I_f M_f| \oprt{Q}(X\lambda \mu) |I_i M_i}}^2. 
\eeq
Note that its unit is $[{\rm e}^2 \cdot ({\rm fm})^{2\lambda}]$ commonly for electric and magnetic $\lambda$th mode.
If both the initial and final states are spherical, 
this can be reduced as
\beq
 B(X\lambda \mu;j_i\rightarrow j_f) = \frac{1}{2j_i+1} \abs{\Braket{j_f|| \oprt{Q}(X\lambda) ||j_i}}^2,
\eeq
by Wigner-Eckart theorem \cite{60Edm}.
In the code \textsc{TOSPEM}, these reduced amplitudes are computed according to Eq.(B.81) and Eq.(B.82) in Ref. \cite{80Ring}.
Namely, for electric modes,
\beqa
\Braket{j_f|| \oprt{Q}(EJ) ||j_i}
={\rm e}\frac{1+(-)^{j_i+j_f+J}}{2} \Braket{u_f\mid r^{J} \mid u_i}  &&  \nonumber \\
w(J,j_f,j_i) (-)^{j_f-1/2} \left( \begin{array}{ccc} j_f &J &j_i \\ -\frac{1}{2} &0 &\frac{1}{2} \end{array}  \right), &&
\eeqa
where
\beq
w(J,j_f,j_i) \equiv \sqrt{\frac{(2J+1) (2j_i +1) (2j_f +1)}{4\pi}}.
\eeq
For magnetic modes, on the other side,
\beqa
\Braket{j_f|| \oprt{Q}(MJ) ||j_i}
=\mu_{\rm N} \frac{1-(-)^{j_i+j_f+J}}{2} \Braket{u_f\mid r^{J-1} \mid u_i} && \nonumber  \\
w(J,j_f,j_i) (-)^{j_f-1/2} \left( \begin{array}{ccc} j_f &J &j_i \\ -\frac{1}{2} &0 &\frac{1}{2} \end{array}  \right) && \nonumber \\
(J-k)\left( \frac{g_s}{2}  -g_l -g_l\frac{k}{J+1} \right), &&
\eeqa
with
\beq
k \equiv \left( j_f + \frac{1}{2} \right) (-)^{l_f+j_f+\frac{1}{2}}  +\left( j_i + \frac{1}{2} \right) (-)^{l_i+j_i+\frac{1}{2}}.
\eeq
Note that the radial integration,
\beq
\Braket{u_f\mid r^N \mid u_i} = \int r^2 dr~u_{n_f,l_j,j_f}(r) r^N u_{n_i,l_i,j_i}(r),
\eeq
is numerically computed in the subroutine ``SPEM'', which provides the solution (A) in the code.
On the other side, this radial integration is approximated
in the Weisskopf's estimate \cite{1951Weiss}, namely, in the solutions (B) and (C) in the code.

\subsection{Weisskopf estimate}
In paper \cite{1951Weiss}, Weisskopf presented an estimate generally for
$EJ$ and $MJ$ transitions of the single proton inside nuclei.
From Weisskopf's estimate \cite{80Ring,1951Weiss}, one can infer that, for the electric mode,
\beq
B_p(E\lambda \mu;I_i\rightarrow I_f)
\cong \frac{1}{4\pi} \left( \frac{3}{\lambda+3} \right)^2 \left( 1.21 A^{1/3} \right)^{2\lambda} ~~\left[ e^2 ({\rm fm})^{2\lambda} \right].
\eeq
For the magnetic mode, by using an approximation as $\left( \frac{g^{(p)}_s}{2} \right)^2 +1 \cong 10$ for proton's spin-gyro-magnetic factor,
\beq
B_p(M\lambda \mu;I_i\rightarrow I_f) \cong \frac{10}{\pi} \left( \frac{3}{\lambda+3} \right)^2 \left( 1.21 A^{1/3} \right)^{2\lambda-2} ~~[\mu_{\rm N}^2({\rm fm})^{2\lambda-2}], 
\eeq
where $\mu_{\rm N}^2 \cong 1.106 \times 10^{-2}$ [${\rm e}^2$fm$^2$].
In this library, instead of Weisskopf's original setting, an approximation of $\left( \frac{g^{(p)}_s}{2} \right)^2 +1  \cong 8.8$ for proton's spin-gyro-magnetic factor is employed.

Pay attention on that, in the neutron case, this estimate cannot apply directly,
because (i) its electric charge is zero, and
(ii) gyro-magnetic factor is different from proton's one.
In this library, I assume that
(i) the neutron has the imaginary electric charge $+e$, as well as
(ii) $g^{(n)}_s=-3.862$ and $g^{(n)}_l=0$.

\section{sample calculation of electric/magnetic transitions}
With the same setting and parameters in section \ref{sec:WERTYEFWQ},
by using the wave functions solved there,
the results of sample calculations for the $^{40}$Ca nucleus are presented.
The $E1$ and $M1$ transitions are computed.
In the code TOSPEM.f90, three method are compared.
Namely, (A) full computation,
(B) Weisskopf estimate, and (C) as the same to (A),
but the radial integration of $B(EJ/MJ)$ is replaced to that in the method (B).

\subsection{E1 transitions}
$E1$ transitions of $0d_{3/2} \longrightarrow 1s_{1/2}$ in $^{40}$Ca.
For the proton,
$B_{p}(E1)=0.236$,
$0.754$, and
$0.754$ e$^2$fm$^2$
in the solutions (A), (B), and (C), respectively.
For the neutron,
$B_{n}(E1)=0.235$,
$0.754$, and
$0.754$ e$^2$fm$^2$
in (A), (B), and (C), respectively.
Note that here I assumed that the neutron has the charge of $+$e commonly to the proton.
Notice that the radial integration of $B(E1)$ is over estimated in (B) and (C),
where the Weisskopf estimate is used.
The output appears as follows.
\begin{lstlisting}[basicstyle=\ttfamily\footnotesize, frame=single]
 Mode of Q(J):           8 for (8:Elec, 9:Mag) with J =           1
  |i>   with (n,l,j*2)=  0       2       3
    <f| with (n,l,j*2)=  1       1       1  (proton)
  |i>   with (n,l,j*2)=     0       2       3
    <f| with (n,l,j*2)=     1       1       1  (neutron)
 <><<><><><><<><><><><<><><><><<><><><><<><><><><<><><><><<><><>
  [A] Suhonen's formulas Eq. (6.23) and (6.24), where
      both the radial and angular integrations are properly computed:
-->   B_{QJ} =|<f|Q(J)|i>|**2 /(2j_i +1) =       0.236420 (proton)
-->                                             0.235329 (neutron)
  ---
  radial integration of <f|Q|i> =  -1.7236413891321813       (prot.)
                                =  -1.7196615102431740       (neut.)
  ---
  angular part of <f|Q_J|i> =  0.56418958360071603   (prot.)
                            =  0.56418958360071603   (neut.)
 <><<><><><><<><><><><<><><><><<><><><><<><><><><<><><><><<><><>
  [B] VS Weisskopf estimate (1) [Phys. Rev. (1951) 83, 1073]:
  Mode of Q(J):   8 for (8:Elec, 9:Mag) with J =           1
-->    B_{QJ} =|<f|Q(J)|i>|**2 /(2j_i +1) =       0.753902 (proton)
-->                                              0.753902 (neutron)
  Same but without A-dep. factor f_J =x^{J}   for EJ mode,
    where x=A^{2/3},        or   f_J =x^{J-1} for MJ mode:
-->           B_{QJ} /f_J      =       0.064458 (proton)
-->                                   0.064458 (neutron) 
 <><<><><><><<><><><><<><><><><<><><><><<><><><><<><><><><<><><>
  [C] VS Weisskopf estimate (2), where
  radial integration is approximated as ~= (3*R**N)/(3 + ilam),
  where R=1.2*ac**(1/3) and N=ilam (ilam-1) for E (M),
  whereas the angular part is computed commonly as in [A]:
-->   B_{QJ} =|<f|Q(J)|i>|**2 /(2j_i +1) =       0.753902 (proton)
-->                                             0.753902 (neutron)
  angular part of <f|Q|i> =  0.564189583600 (prot.)
                          =  0.564189583600 (neut.)
  ---
  radial part of <f|Q|i>, approximated as
  ~= (3*R**N)/(3 + ilam) =   3.07795670   (prot.)
                         =   3.07795670   (neut.)
\end{lstlisting}

\subsection{M1 transitions}
$M1$ transitions of $0f_{7/2} \longrightarrow 0f_{5/2}$ in $^{40}$Ca.
For the proton,
$B_{p}(M1)=2.145$,
$1.576$, and
$1.210~\mu_{\rm N}^2$
in the solutions (A), (B), and (C), respectively.
For the neutron,
$B_{n}(M1)=1.496$,
$0.655$, and
$0.842~\mu_{\rm N}^2$
in (A), (B), and (C), respectively.
Note that different $g$ factors are used between protons and neutrons.
The output appears as follows.
\begin{lstlisting}[basicstyle=\ttfamily\footnotesize, frame=single]
Mode of Q(J):    9 for (8:Elec, 9:Mag) with J =  1
 
 |i>   with (n,l,j*2)=           0           3           7
   <f| with (n,l,j*2)=           0           3           5  (proton)
 |i>   with (n,l,j*2)=              0           3           7
   <f| with (n,l,j*2)=              0           3           5  (neutron)
 
 <><<><><><><<><><><><<><><><><<><><><><<><><><><<><><><><<><><>
  [A] Suhonen's formulas Eq. (6.23) and (6.24), where
      both the radial and angular integrations are properly computed:
-->         B_{QJ} =|<f|Q(J)|i>|**2 /(2j_i +1) =       2.145306 (proton)
-->                                                   1.496479 (neutron)
  ---
  radial integration of <f|Q|i> =  0.99848891462729705       (prot.)
                                =  0.99959266730175311       (neut.)
  ---
  angular part of <f|Q_J|i> =  -4.1490278177880677       (prot.)
                          =   3.4614436177185230       (neut.)
  ---
  angular part of <f|Q_J|i>, but the factor,
  sqrt((2J+1)/4pi) is eliminated: =  -8.4916219551138372       (prot.)
                                  =   7.0843754034595596       (neut.)
 
 <><<><><><><<><><><><<><><><><<><><><><<><><><><<><><><><<><><>
  [B] VS Weisskopf estimate (1) [Phys. Rev. (1951) 83, 1073]:
  Mode of Q(J):           9 for (8:Elec, 9:Mag) with J =           1
-->           B_{QJ} =|<f|Q(J)|i>|**2 /(2j_i +1) =       1.575786 (proton)
-->                                                     0.655243 (neutron)
  Same but without A-dep. factor f_J =x^{J}   for EJ mode,
    where x=A^{2/3},        or   f_J =x^{J-1} for MJ mode:
-->           B_{QJ} /f_J      =       1.575786 (proton)
-->                                   0.655243 (neutron)
 
 <><<><><><><<><><><><<><><><><<><><><><<><><><><<><><><><<><><>
  [C] VS Weisskopf estimate (2), where
  radial integration is approximated as ~= (3*R**N)/(3 + ilam),
  where R=1.2*ac**(1/3) and N=ilam (ilam-1) for E (M),
  whereas the angular part is computed commonly as in [A]:
-->           B_{QJ} =|<f|Q(J)|i>|**2 /(2j_i +1) =       1.210390 (proton)
-->                                                     0.842456 (neutron)
  angular part of <f|Q|i> =  -4.1490278177880677       (prot.)
                          =   3.4614436177185230       (neut.)
  ---
  radial part of <f|Q|i>, approximated as
  ~= (3*R**N)/(3 + ilam) =  0.75000000000000000       (prot.)
                         =  0.75000000000000000       (neut.)
\end{lstlisting}
Notice that, in the $M1$ solution (A), its radial integration is approximately one.
This is natural because, for the upper and lower levels of LS-split partners,
the radial integration reads $\int r^2 u_{l_>}(r) u_{l_<}(r) dr \cong 1$,
when the LS potential does not strongly affect their radial profiles.

\section{historical memos}
\subsection{memo on 2022/11/21}
Today I have finished to build up the core part of this \textsc{TOSPEM}.
I noticed that, in Ref. \cite{1951Weiss} by Weisskopf, a rough estimate for proton's
gyro-magnetic factor was used.

\subsection{memo on 2023/02/09}
Today I have examined the consistency of $M1$ strength with that by my \textsc{CPNM1} code \cite{2019OP}.
Unfortunately, there remain some errors:
the angular part of $B(M1)$ is not equal to that by the \textsc{CPNM1} code.
Note that \textsc{TOSPEM} and \textsc{CPNM1} are not on the equal documentations:
the former one is based on Dr. Nanashino's note, whereas the later one is based on Oishi's note for $M1$ transitions.
Also, the \textsc{CPNM1} result was checked as consistent to the Zagreb formula,
namely $m_0(M1) = 2N_{S=1}$ in Ref. \cite{2019OP}.
Finally, I consider that there can be bug(s) in the \textsc{TOSPEM} side.

\subsection{memo on 2023/03/22}
The old angular part of $B_{MJ}$ was eliminated.
Then, the new angular part is installed as the copy of ``dbmmag\_*\_RS'' in the \textsc{CPNM1} code.
This code is based on the textbook by Ring and Schuck \cite{80Ring}, as well as
``xljj'' and ``xsjj'' in Oishi's note.
Their results were checked as consistent to the Zagreb formula \cite{2019OP}.

\chapter{Library-02: RESONA} \label{Ch_RESONA}
Purpose of library RESONA is to solve the resonant eigenstates of spherical Sch\"{o}dinger equations.
On 2023/March/09th, I had started to build up this library.
Then, the four different solvers, (i) the graphic fitting of energy stabilization, (ii) time-dependent (TD) calculation, (iii) scattering phase-shift calculation, and (iv) complex scaling method \cite{2014Myo_rev,2022Myo_rev}
were implemented until 2023/April/10th.
A rough consistency between these methods was confirmed by solving a common problem, namely, the resonance state of $^{16}$O$+n(d_{3/2})$ as described in the following sections.
In parallel, the four solvers are based on the independent logics and algorithms,
including different types of errors.
Consequently, their results show finite deviations.

\section{setting for benchmark calculations}
In this library, a spherical Sch\"{o}dinger equation is solved for benchmark.
That reads
\beq
\hat{H} \psi(r) =
\left[ -\frac{\hbar^2}{2\mu} \frac{d^2}{dr^2} +\frac{\hbar^2}{2\mu} \frac{l(l+1)}{r^2} + V(r)  \right] \psi(r) = \epsilon \psi(r), \label{eq:TDSCPS}
\eeq
where $\epsilon = E_r -i\Gamma/2$ for the unbound, resonant state \cite{89Kuku}.
I assume the $^{16}$O+$n$ system, and thus, $\mu = m_n m_c /(m_n + m_c)$,
where $m_c$ ($m_n$) is the core-nucleus (neutron) mass.
The single-particle potential $V(r)$ is determined as the Woods-Saxon plus spin-orbit potential.
That is given as
\beqa
&& V(r) = V_{WS}(r) = V_0 f(r) + U_{ls} (\bi{l} \cdot \bi{s}) \frac{1}{r} \frac{df(r)}{dr}, \nonumber  \\
&& f(r) = \frac{1}{1 + e^{(r-R_0)/a_0}}, \label{eq:cp_WS}
\eeqa
where $R_0=r_0\cdot A_{core}^{1/3}=r_0\cdot 16^{1/3}$, and $f(r)$ is the standard Fermi profile.
In this library, I fix the parameters as $V_0=-53.2$ MeV,
$U_{ls}=22.1$ MeV$\cdot$fm$^2$,
$r_0 =1.25$ fm, and $a_0 =0.65$ fm.
Note also that, if the particle was proton, the Coulomb potential $V_{Coul}(r)$ of an uniformly charged sphere with radius $R_0$ should be employed for the core-proton potential: $V(r)= V_{WS}(r) +V_{Coul}(r)$.

The potential (\ref{eq:cp_WS}) has the single-neutron bound states with
$E_n(0s_{1/2})=-31.970$ MeV,
$(0p_{3/2})=-17.846$ MeV,
$(0p_{1/2})=-14.594$ MeV,
$(0d_{5/2})=- 4.143$ MeV, and
$(1s_{1/2})=- 3.275$ MeV.
Then, in the $0d_{3/2}$ channel, the first resonant state could exist.
Note that this neutron resonance appears in some excited state, not in the ground state.
For numerical calculations, in addition to physical parameters,
the cutoff energy $E_{\rm cut}$ and radial-box size $r_{\rm max}$ are
employed in the source codes.

%%%%%%%%%%%%%%%%%%%%%%%%%%%%%%%%%%%%%%%%%%%%%%%%%%%%%%%%%%%%%%%%%%%%%%
\begin{figure}[tb] \begin{center}
\includegraphics[width = 0.4\hsize]{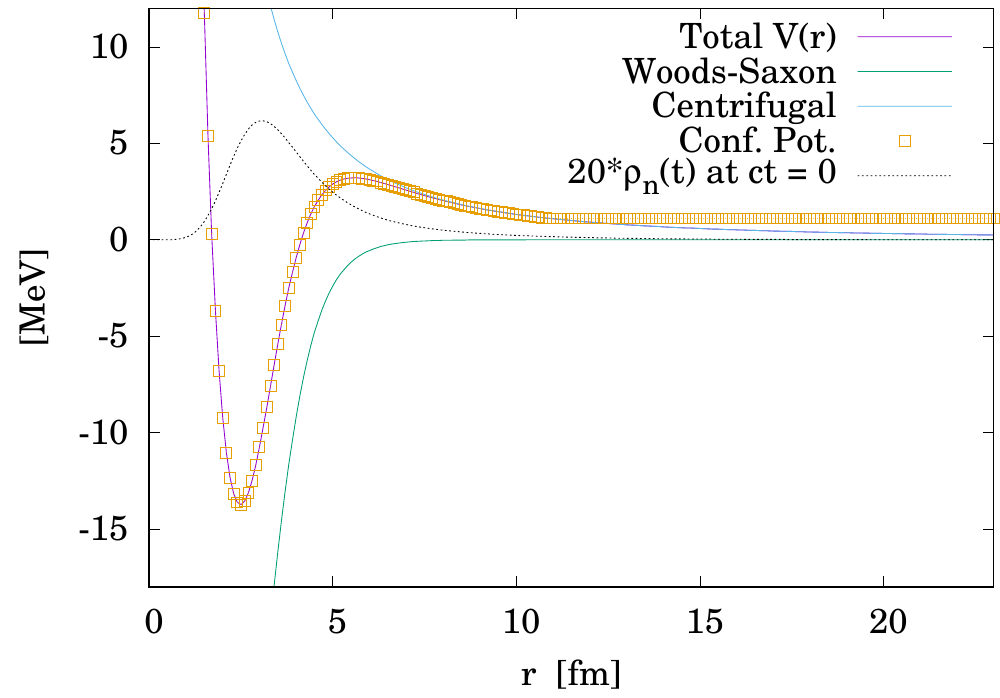}
\includegraphics[width = 0.4\hsize]{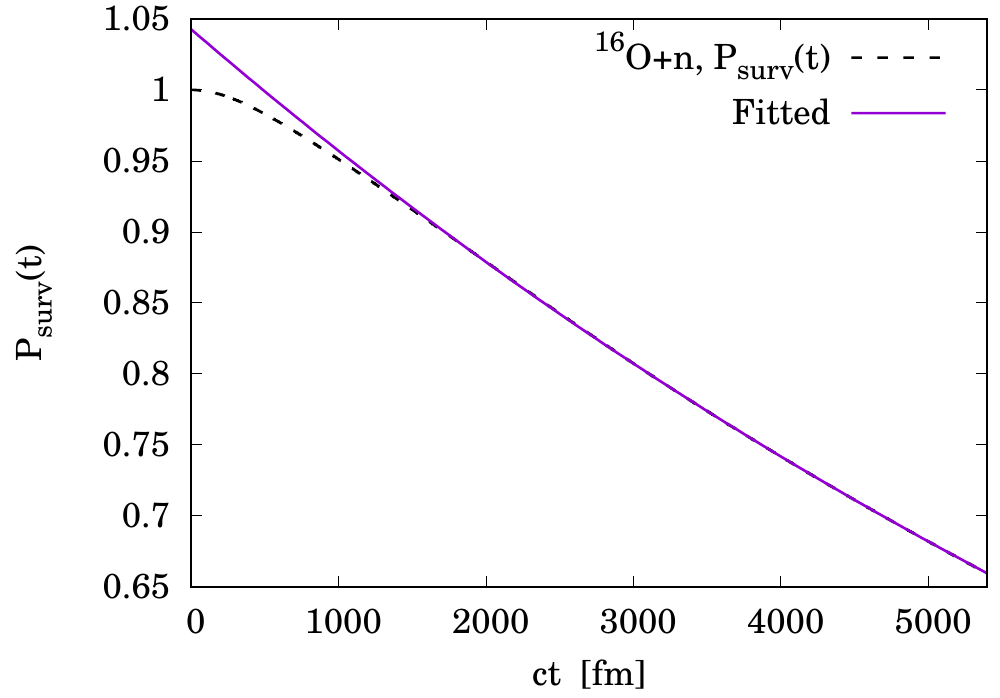}
\caption{(Left)
Single-particle potential for $^{16}$O+$n(d_{3/2})$.
The neutron density of the initial state, $\rho_{n}(t=0)$, in the time-dependent calculation is also displayed.
(Right)
Survival probability $P_{\rm surv}(t)$ obtained with the time-dependent method.
The fitted function, $C(t)=\exp (-ct \Gamma /\hbar c) +0.0427875$ with $\Gamma = 0.0176381$ MeV,
is also presented.
}
\label{fig:W1_TDM}
\end{center} \end{figure}
%%%%%%%%%%%%%%%%%%%%%%%%%%%%%%%%%%%%%%%%%%%%%%%%%%%%%%%%%%%%%%%%%%%%%%

\section{time-dependent method}
Result: $E_r=420.9$ keV, $\Gamma=17.6$ keV, for $^{16}$O$+n(d_{3/2})$.

Time-dependent picture enables one to interpret the nucleon's resonance
as the radioactive process with quantum-tunneling effect \cite{89Kuku,87Gur,88Gur,94Serot,94Car}.
In this picture, the resonance energy $E_r$ coincides the mean energy release (Q value),
whereas the width $\Gamma$ is evaluated from the lifetime.

The source code ``SCPSTD.f90'' is prepared for time-dependent calculations.
In the main program ``main\_clock'', the subroutine ``CN\_TD'' executes the
time-dependent calculations.
Physical and numerical parameters are determined in the module ``mdl\_001\_setting'', except the Woods-Saxon parameters determined in  ``mdl\_003\_subroutines''.
This code solves the Schr\"{o}dinger equation with the Numerov method,
where its details are summarized in Appendix \ref{Ap_Numerov}.
The box size is fixed as $R_{\rm max}=240$ fm to obtain the following results.

Figure \ref{fig:W1_TDM} shows the core-neutron potential in the $d_{3/2}$ ($l=2,~j=3/2$) channel.
Notice that the total potential $V(r)$ has a barrier from the combination of
Woods-Saxon and centrifugal potentials.
For time-dependent calculations,
the initial state is determined within the confining potential as shown in Fig. \ref{fig:W1_TDM}.
Namely, I simply assume the confining wall for $r \ge 11$ fm.
One can read that, at $t=0$, this initial state is well confined inside the potential
barrier around $r\cong 10$ fm.
The $1n$ energy of this initial state is obtained as
$\Braket{\phi(0) \mid \hat{H} \mid \phi(0)}=420.9$ keV,
where $\hat{H}$ indicates the original single-particle Hamiltonian given in Eq. (\ref{eq:TDSCPS}).
Note that, by using the eigenstates of $\hat{H}$, namely $\left\{ \psi_n \right\}$ with
$\hat{H} \ket{\psi_n} =E_n \ket{\psi_n}$,
the initial state is expanded as
\beq
\ket{\phi(t=0)} = \sum_{n} \alpha_n \ket{\psi_n}
\eeq
in numerical calculations, but only including the $d_{3/2}$ channel since the Hamiltonian is spherical.
Then, the time development is obtained as
\beq
\ket{\phi(t)} = e^{-it\frac{\hat{H}}{\hbar}} \ket{\phi(0)} = \sum_{n} e^{-it\frac{E_n}{\hbar}} \alpha_n \ket{\psi_n}.
\eeq
The survival probability, which is shown in Fig. \ref{fig:W1_TDM}, is evaluated as
\beq
P_{\rm surv}(t) \equiv \abs{\Braket{\phi(t) \mid \phi(0)} }^2 \cong P_{\rm surv}(0) \cdot e^{-t/\tau},
\eeq
where $P_{\rm surv}(0) \equiv 1$ (initial normalization).
Namely, the calculated result can be well approximated as an exponential-decaying process.
As the result of fitting, the $1n$-resonance width is obtained as $\Gamma \cong 17.6$ keV from $\tau=\hbar / \Gamma$.
%This behaviour is consistent to the quantum-tunneling picture.

%%%%%%%%%%%%%%%%%%%%%%%%%%%%%%%%%%%%%%%%%%%%%%%%%%%%%%%%%%%%%%%%%%%%%%
\begin{figure}[tb] \begin{center}
\includegraphics[width = 0.4\hsize]{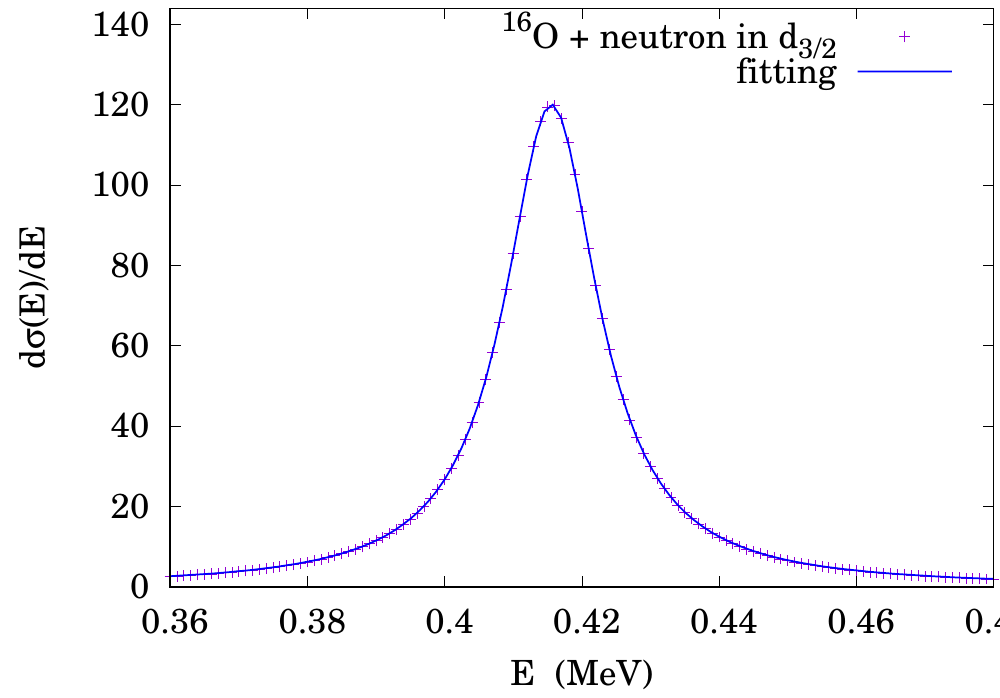}
\caption{
Derivative of scattering phase shift, $d\sigma /dE$, obtained in
the benchmark case of $^{16}$O+$n(d_{3/2})$.
The Cauchy-Lorentz profile after fitting is also presented.
}
\label{fig:W2_SCAT}
\end{center} \end{figure}
%%%%%%%%%%%%%%%%%%%%%%%%%%%%%%%%%%%%%%%%%%%%%%%%%%%%%%%%%%%%%%%%%%%%%%

\section{scattering phase-shift calculation}
Result: $E_r=415.5$ keV, $\Gamma=16.6$ keV, for $^{16}$O$+n(d_{3/2})$.

Mathematical details of phase-shift calculations are summarized in Appendix \ref{Ap_Scat_2body}.
By solving the quantum-mechanical two-body scattering problem in the spherical case \cite{07Sasakawa,89Kuku},
one obtains the scattered wave function $\Psi(kr)$ as
\beq
\Psi(kr) \cong \Phi(kr +\sigma(E))~~~(r\longrightarrow \infty),
\eeq
in the asymptotic region.
Here $\Phi(kr)$ is the asymptotic solution when the potential of interest was absent,
whereas $\sigma(E)$ indicates the phase shift due to that potential.
When the resonant state exists, the derivative $d\sigma(E)/dE$ often shows the corresponding Cauchy-Lorentz profile.

The phase shift of scattering problem of Eq. (\ref{eq:TDSCPS}) is computed within the code ``SCPSTD.f90'', which is the same code for the time-dependent calculations.
In the main program ``main\_clock'', the subroutine ``CN\_SC'' executes the
phase-shift calculations.
For computing the phase shift $\sigma(E)$,
the open-source code ``COULCC'' by Thompson and Barnett is copied and utilized \cite{1985Thompson_COULCC}.

Figure \ref{fig:W2_SCAT} shows the phase-shift $\sigma(E)$ calculated in the $d_{3/2}$ ($l=2,~j=3/2$) channel.
There, the obtained $d\sigma /dE$ distribution is fitted by a Cauchy-Lorentz (Breit-Wigner) profile:
\beq
F(E) = \frac{1}{\pi} \frac{\Gamma/2}{(E-E_r)^2 +\Gamma^2/4}.
\eeq
As the result, this $^{16}$O+$n$ system with the potential given in Eq. (\ref{eq:cp_WS}) has one resonant state in the $d_{3/2}$ channel at $E_r=415.5$ keV with the width $\Gamma=16.6$ keV.

%%%%%%%%%%%%%%%%%%%%%%%%%%%%%%%%%%%%%%%%%%%%%%%%%%%%%%%%%%%%%%%%%%%%%%
\begin{figure}[tb] \begin{center}
\includegraphics[width = 0.4\hsize]{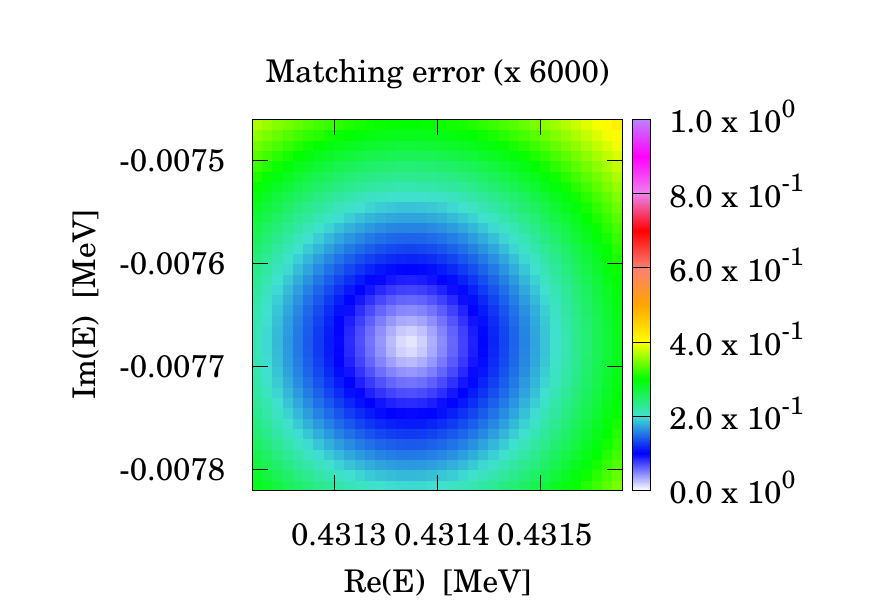}%{./Libra_02_RESONA/plocs_16O_n_d32.pdf}
\caption{Matching error of the complex-scaled Runge-Kutta calculation for $^{16}$O$+n(d_{3/2})$.} \label{fig:W3_YVNFFX}
\end{center} \end{figure}
%%%%%%%%%%%%%%%%%%%%%%%%%%%%%%%%%%%%%%%%%%%%%%%%%%%%%%%%%%%%%%%%%%%%%%

\section{complex-scaling method}
Result: $E=431.4$ keV, $\Gamma=15.4$ keV, for $^{16}$O$+n(d_{3/2})$.

In the source code ``SCHRCS.f90'',
the complex-scaled Runge-Kutta (CSRK4) method is
utilized to obtain the complex eigen energy,
$\epsilon=E_r -i\Gamma/2$, of Eq. (\ref{eq:TDSCPS}).
See Appendix \ref{Ap_RK4} for details of the complex-scaled Runge-Kutta method.
By introducing the new symbols, $A(r)\equiv r\cdot \psi(r)$ and $B(r) \equiv \frac{dA}{dr}$,
the original Schr\"{o}dinger equation (\ref{eq:TDSCPS}) transforms as
\beq
\frac{d}{dz}
\left( \begin{array}{c} A(z) \\ B(z) \end{array} \right)
=
\left( \begin{array}{cc} 0 &1 \\ G(z,\epsilon) &0 \end{array} \right)
\left( \begin{array}{c} A(z) \\ B(z) \end{array} \right),
\eeq
with $z=e^{i\theta} r$, where $\alpha$ is the complex-scaling angle.
For solving the typical resonance of $\epsilon = E_r -i\Gamma/2$,
one needs to set $\theta > \frac{1}{2} \arctan \left( \frac{\Gamma}{2E} \right)$
in numerical calculations \cite{1997Myo}.
Note that $G(z,\epsilon)$ includes the single-particle potential, $V(z)$.
By using this Runge-Kutta method with complex variables,
the forward and backward solutions are obtained as
$\left\{A_F(r), B_F(r) \right\}$ and 
$\left\{A_B(r), B_B(r) \right\}$, respectively.
Then, as one numerical technique, the error of matching, $X(\epsilon, r_m)$,
is evaluated as
\beq
X(\epsilon, r_m) \equiv \sqrt{W(\epsilon, r_m) \cdot W(\epsilon, 2r_m)},~~~
W(\epsilon, r_m)=\abs{B_F A_B - A_F B_B}_{r=r_m}
\eeq
where $r_m=2.6$ fm in this sample calculation.
The result is displayed in Fig. \ref{fig:W3_YVNFFX}.
One can find that the matching error has the minimum around $E_r=431.4$ keV and $\Gamma/2=7.7$ keV.
Note that there can be a better option for evaluating the matching error than the current one.

%%%%%%%%%%%%%%%%%%%%%%%%%%%%%%%%%%%%%%%%%%%%%%%%%%%%%%%%%%%%%%%%%%%%%%
\begin{figure}[tb] \begin{center}
\includegraphics[width = 0.4\hsize]{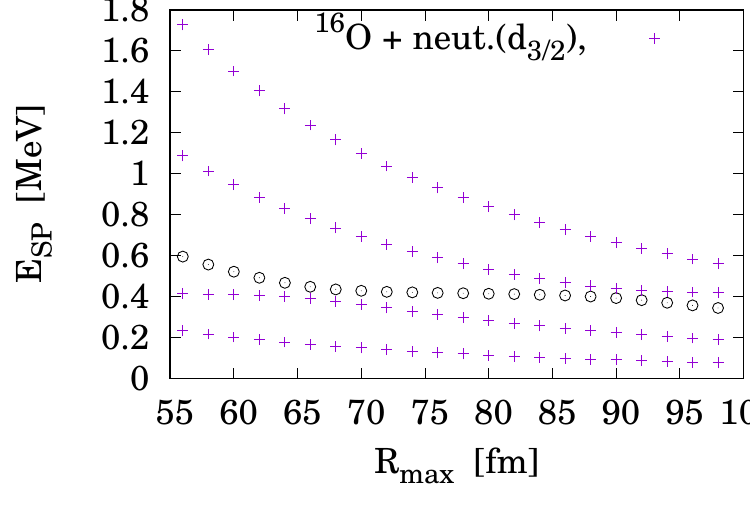}%{./Libra_02_RESONA/plst1_16O_n_d32.pdf}
\includegraphics[width = 0.4\hsize]{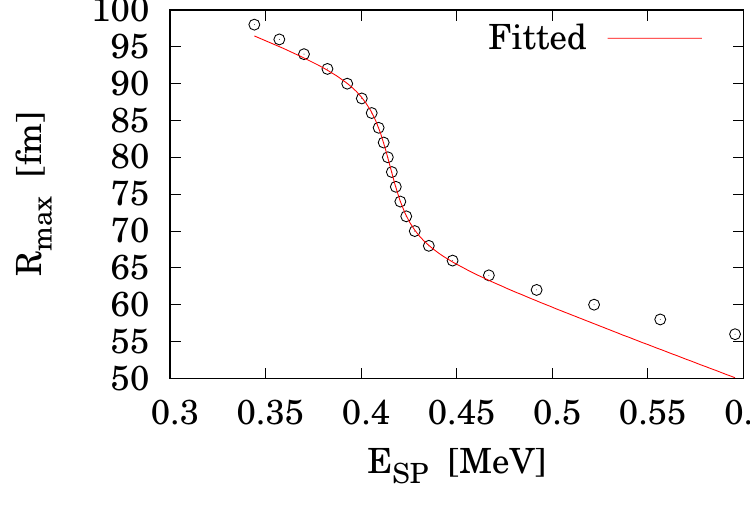}%{./Libra_02_RESONA/plst2_16O_n_d32.pdf}
\caption{(Left) Single-particle energies for $^{16}$O$+n(d_{3/2})$ obtained
by changing the radial-box size.
(Right)
Inverse-function fitting result for $R_{\rm max}(E_{\rm sp})$.
}
\label{fig:W3_STAB}
\end{center} \end{figure}
%%%%%%%%%%%%%%%%%%%%%%%%%%%%%%%%%%%%%%%%%%%%%%%%%%%%%%%%%%%%%%%%%%%%%%

\section{energy stabilization method}
Result: $E_r=415.1$ keV, $\Gamma=16.7$ keV, for $^{16}$O$+n(d_{3/2})$, obtained
with ``SPES.f90'' (source code) and ``Sequence.sh'' (BASH script).

As one empirical low, for a resonant state with the complex eigen energy
$\epsilon=E_r -i\Gamma/2$, numerical solutions are often stabilized around $E_r$
even when the model space is modified \cite{1970Hazi,2011Ghoshal}.
In the present case of $^{16}$O$+n(d_{3/2})$,
the radial-box size $r_{\rm max}$ can be utilized to confirm this stability.
Indeed, as shown in Fig. \ref{fig:W3_STAB}, numerical solutions of discretized-continuum energies
become stable around $E\cong 0.4$ MeV.
The inverse-function fitting enables one to evaluate the resonance width $\Gamma$.
Namely, by assuming a smooth background,
\beq
\frac{dR_{\rm max}}{dE} \cong c_1 \frac{\Gamma/2}{(E-E_r)^2 +(\Gamma/2)^2} +c_2.
\eeq
Or equivalently,
\beq
R_{\rm max}(E) \cong c_1 \arctan \left[ \frac{E-E_r}{\Gamma/2}  \right] +c_2 E +R_0.
\eeq
Figure \ref{fig:W3_STAB} displays the result of fitting.
The resonance energy and width are evaluated as $E_r=415.1$ keV and $\Gamma=16.7$ keV, respectively.

For computing the eigen energies above the threshold by changing $r_{\rm max}$, I used the BASH-script file,
``Sequence.sh'', as following.
\begin{lstlisting}[basicstyle=\ttfamily\footnotesize, frame=single]
#!/bin/sh
l=700000
n=560
z=20
rm  OUTRU*.txt  Rmax_and_En*.txt
while [ $n -ne 1000 ]
do
  echo "(N=${n})-operation"
  sed "3 s/=.../=$n/" SPES.f90 > Main.f90
  gfortran  Main.f90  -o  run.out
  b=`expr $n + $l`
  time ./run.out > OUTRUN_${b}.txt
  mv  Rmax_and_Energies.txt  Rmax_and_Enes_${b}.txt
  n=`expr $n + $z`
  echo "----------------------------------------------------------"
done
cat  Rmax_and_Enes_*.txt  >  Fine.dat
\end{lstlisting}
Here the original f90 file, ``SPES.f90'', includes the sentences,
\begin{lstlisting}[basicstyle=\ttfamily\footnotesize, frame=single]
module mdl_001_setting                                                !L0001
   implicit none                                                      !L0002
   integer, parameter :: nrmax =xxx                                   !L0003
   double precision, parameter :: dr = 0.1d0                          !L0004
   double precision, parameter :: rmax = dr*(dble(nrmax) +1.d-11)     !L0005
\end{lstlisting}
where the 3rd line is modified in each iteration.

%\include{end}

%\input{ch3.tex}
%\input{ch0.tex}
%---
\appendix
\pagestyle{fancy}
  \fancyhead{} %reset headings
  \fancyhead[LE,RO]{\leftmark}
  \renewcommand{\chaptermark}[1]{\markboth{Appendix~\thechapter\; \; #1}{}}
  \cfoot{\thepage}

\chapter{Runge-Kutta method} \label{Ap_RK4}

\section{radial wave functions}
The code SCHRCS.f90 solves both the bound and resonant states with the (complex-scaled) Runge-Kutta method of spherical Schr\"{o}dinger equation.
That is,
\beq
\left[ -\frac{\hbar^2}{2\mu} \nabla_{\bir}^2 +V(r) \right] \psi(\bir) = E \psi(\bir).
\eeq
The spherical solution generally reads
\beq
\psi(\bir) = \frac{A(r)}{r} Y_{lm}(\theta,\phi).
\eeq
Thus, by using $v(r) \equiv 2\mu V(r)/\hbar^2$ and $\epsilon \equiv 2\mu E/\hbar^2$,
the equation changes as
\beqa
\left[ -\left( r^{-1}\frac{d^2}{dr^2}r -\frac{l(l+1)}{r^2} \right) +v(r)-\epsilon \right] \frac{A(r)}{r} &=& 0 \nonumber \\
\longrightarrow \left[ \frac{d^2}{dr^2}  -\frac{l(l+1)}{r^2} -v(r) +\epsilon \right] A(r) &=& 0 \nonumber  \\
\frac{d^2}{dr^2} A(r) = G_E(r) A(r), \label{eq:VHSUYTE} &&
\eeqa
where
\beq
G_E(r) \equiv \frac{2\mu V(r)}{\hbar^2} +\frac{l(l+1)}{r^2} -\frac{2\mu E}{\hbar^2}.
\eeq
By using $B(r) \equiv \frac{dA}{dr}$, Eq.(\ref{eq:VHSUYTE}) becomes
\beq
\frac{dA}{dr} = B(r),~~~~
\frac{dB}{dr} = G_E(r) A(r).
\eeq
Or equivalently,
\beq
\frac{d}{dr}
\left( \begin{array}{c} A(r) \\ B(r) \end{array} \right)
=
\left( \begin{array}{cc} 0 &1 \\ G_E(r) &0 \end{array} \right)
\left( \begin{array}{c} A(r) \\ B(r) \end{array} \right).
\eeq
This equation can be solved with the Runge-Kutta method when their asymptotic forms are given
for starting condition.
In our case,
\beqa
A(r\rightarrow 0) &\cong &  r^{l+1}, \\
A'(r\rightarrow 0) &\cong & (l+1) r^{l+1}, \nonumber
\eeqa
as well as, when $V(r\rightarrow \infty)=0$,
\beqa
A(r\rightarrow \infty) &\cong & re^{-kr}, \\
A'(r\rightarrow \infty) &\cong & (1-kr)e^{-kr}, \nonumber
\eeqa
where $k=\sqrt{-2\mu E} /\hbar$ for the bound state with $E<0$.
Note also that these asymptotic forms help us to infer whether the numerical routines
are correctly installed: if there is a bug, the output wave functions do not behave like these asymptotic forms.

\section{implementation of Runge-Kutta method}
Equation of interest here is given as
\beq
\frac{d}{dr}
\left( \begin{array}{c} A(r) \\ B(r) \end{array} \right)
=
\left( \begin{array}{cc} a(r) &b(r) \\ c(r) &d(r) \end{array} \right)
\left( \begin{array}{c} A(r) \\ B(r) \end{array} \right).
\eeq
We assume that $A_0 =A(r_0)$ and $B_0 =B(r_0)$ are already known.
Runge-Kutta method gives the numerical solution in mesh points $r_{n+1}=r_n+D$ as \cite{1989Atkinson}
\beqa
&& A_{n+1} = A_n +\frac{D}{6} \left( u_1 +2u_2 +2u_3 +u_4  \right), \nonumber  \\
&& B_{n+1} = B_n +\frac{D}{6} \left( v_1 +2v_2 +2v_3 +v_4  \right).
\eeqa
Here the 1st to 4th-step elements read
\beqa
&&u_1 = \left[ a(s_1)       A_n                           + b(s_1)B_n \right], \nonumber \\
&&u_2 = \left[ a(s_2)\left( A_n +\frac{D}{2}u_1 \right)   + b(s_2)\left( B_n +\frac{D}{2}v_1 \right) \right], \nonumber \\
&&u_3 = \left[ a(s_3)\left( A_n +\frac{D}{2}u_2 \right)   + b(s_3)\left( B_n +\frac{D}{2}v_2 \right) \right], \nonumber \\
&&u_4 = \left[ a(s_4)\left( A_n +D u_3 \right)            + b(s_4)\left( B_n +D v_3          \right) \right],
\eeqa
and
\beqa
&&v_1 = \left[ c(s_1)       A_n                           + d(s_1)B_n \right], \nonumber \\
&&v_2 = \left[ c(s_2)\left( A_n +\frac{D}{2}u_1 \right)   + d(s_2)\left( B_n +\frac{D}{2}v_1 \right) \right], \nonumber \\
&&v_3 = \left[ c(s_3)\left( A_n +\frac{D}{2}u_2 \right)   + d(s_3)\left( B_n +\frac{D}{2}v_2 \right) \right], \nonumber \\
&&v_4 = \left[ c(s_4)\left( A_n +D u_3 \right)            + d(s_4)\left( B_n +D v_3          \right) \right],
\eeqa
within the coordinates, $s_1=r_n$, $s_2=s_3=r_n +\frac{D}{2}$, and $s_4=r_{n+1}$.
This Runge-Kutta method can be used even if the coordinates and functions are complex.

For the continuous condition, one should 
repeat calculations by changing the energy $E$ until satisfying that
\beqa
&&f_F(r_m) = f_B(r_m),~~~f(r)=\frac{A(r)}{r}, \\
&&f'_F(r_m) = f'_B(r_m),~~~f'(r)=\frac{B(r)}{r}-\frac{A(r)}{r^2},
\eeqa
at the matching point $r_m$, where $f_F(r)$ and $f_B(r)$ are the forward and backward solutions, respectively.

\section{complex-scaled version}
For solving the resonant state with the complex-eigen energy, $\epsilon=E-i\Gamma/2$,
the complex-scaling is utilized.
Thus, the above Runge-Kutta equation is modified as
\beq
\frac{d}{dz}
\left( \begin{array}{c} A(z) \\ B(z) \end{array} \right)
=
\left( \begin{array}{cc} a(z) &b(z) \\ c(z) &d(z) \end{array} \right)
\left( \begin{array}{c} A(z) \\ B(z) \end{array} \right)
=
\left( \begin{array}{cc} 0 &1 \\ G(z,\epsilon) &0 \end{array} \right)
\left( \begin{array}{c} A(z) \\ B(z) \end{array} \right),
\eeq
where $z=e^{i\alpha} r$.
The complex eigenvalue is determined so as to satisfy the matching condition commonly to the bound-state case.

\chapter{Numerov method} \label{Ap_Numerov}
This is the copy of Appendix A in T. Oishi's doctoral thesis \cite{2014Oishi_DT}.

Numerov provided one numerical method to solve an ordinary differential equation in which only the zeroth and the second order terms are included, such as 
\beq
 \left[ \frac{d^2}{dx^2} + f(x) \right] U(x) = 0, \label{eq:Num0} 
\eeq 
where $f(x)$ is an arbitrary source function.
The solution, $U(x)$, is sampled at equidistant points $x_n,(n=0 \sim N)$ 
where the distance between two sampling points is defined as $a$. 
With this method, starting from the solution values at 
two consecutive sampling points, 
namely $U_0 \equiv U(x_0)$ and $U_1 \equiv U(x_1)$, 
we can calculate the remaining solution values as 
\beq
 U_{n+2} = \frac{ (2-5a^2f_{n+1}/6)U_{n+1} - (1+a^2f_{n})U_{n} }
                { 1+a^2f_{n+2}/12 } + \mathcal{O}(a^6), \label{eq:Num1} 
\eeq
where we neglect $\mathcal{O}(a^6)$. 
The derivation of Eq.(\ref{eq:Num1}) is based on the discrete Taylor expansion 
for $U(x)$ until the fifth order. 
Considering the two sampling points, $x_{n-1}=x_n-a$ and $x_{n+1}=x_n+a$, 
Taylor expansions are given as 
\beqa
 U_{n+1} &\equiv& U(x_n+a) \nonumber \\
 &=& U_n+aU'_n+\frac{a^2}{2!}U''_n
        +\frac{a^3}{3!}U^{(3)}_n+\frac{a^4}{4!}U^{(4)}_n
        +\frac{a^5}{5!}U^{(5)}_n+\mathcal{O}(a^6), \\
 U_{n-1} &\equiv& U(x_n-a) \nonumber \\
 &=& U_n-aU'_n+\frac{a^2}{2!}U''_n
        -\frac{a^3}{3!}U^{(3)}_n+\frac{a^4}{4!}U^{(4)}_n
        -\frac{a^5}{5!}U^{(5)}_n+\mathcal{O}(a^6), 
\eeqa
where $U^{(m)}_n \equiv \left. d^mU(x)/dx^m \right|_{x=x_n}$. 
The sum of these two equations gives 
\beq
  U_{n-1} + U_{n+1} = 2U_n + a^2U''_n + \frac{a^4}{12}U^{(4)}_n + \mathcal{O}(a^6). 
\eeq
Solving this equation for $a^2U''_n$ leads to 
\beq
 -a^2U''_n = 2U_n - U_{n-1} - U_{n+1} + \frac{a^4}{12}U^{(4)}_n + \mathcal{O}(a^6). 
 \label{eq:Num4} 
\eeq
In this equation, we can replace $U''_n$ to $-f_n U_n$ because of Eq.(\ref{eq:Num0}). 
Similarly, for the fourth term in the right hand side, we can use 
\beq
 U^{(4)}(x) = \frac{d^2}{dx^2}[-f(x)U(x)]. 
\eeq
The numerical definition of the second derivative is given as 
the second order difference quotient, that is 
\beq
 \frac{d^2}{dx^2}[-f(x)U(x)] \Rightarrow 
 -\frac{f_{n-1}U_{n-1} - 2f_nU_n +f_{n+1}U_{n+1}}{a^2}. 
\eeq
After these replacements, Eq.(\ref{eq:Num4}) is transformed as 
\beq
 a^2f_nU_n = 
 2U_n - U_{n-1} - U_{n+1} - 
 \frac{a^4}{12} \frac{f_{n-1}U_{n-1} - 2f_nU_n +f_{n+1}U_{n+1}}{a^2} + 
 \mathcal{O}(a^6). 
\eeq
Finally, we solve this equation for $U_{n+1}$ to get 
\beq
 U_{n+1} = \frac{ (2-5a^2f_n/6)U_n - (1+a^2f_{n-1})U_{n-1} }
                { 1+a^2f_{n+1}/12 } + \mathcal{O}(a^6), 
\eeq
which is equivalent to Eq.(\ref{eq:Num1}).

\chapter{Two-body scattering with spherical potential} \label{Ap_Scat_2body}
This is the copy of Appendix D in T. Oishi's doctoral thesis \cite{2014Oishi_DT}.

Our goal here is to derive the fitting formula for the phase-shift 
of two-body scattering problems. 
For simplicity, we assume that the potential between 
two particles is spherical. 
For quantum resonances in two-body systems, one can usually 
solve the asymptotic waves analytically. 
The phase shift and its derivative can be computed by using these asymptotic 
waves, where it indicates the pole(s) of the S-matrix for the resonance. 
Even if one is interested in the scattering problem with three 
or more particles, 
it is often necessary to solve the partial two-body systems in order to, 
{\it e.g.} prepare the fine two-body interactions. 

\section{solutions in asymptotic region}
Assuming the relative wave function as 
$\phi_{ljm}(\bir,\bis) = R_{lj}(r) \mathcal{Y}_{ljm}(\ubir,\bis)$, 
the radial equation of this problem reads 
\beq
  \left[ -\frac{\hbar^2}{2\mu}\left\{ \frac{d^2}{dr^2} - \frac{l(l+1)}{r^2} \right\} + V_{lj}(r) - E \right] U_{lj} (r,E) = 0, 
\eeq
where we defined $U_{lj}(r,E) \equiv rR_{lj}(r)$ from the radial wave function. 
The relative energy, $E$, for the scattering problem satisfies 
\beq
 E > \lim_{r \rightarrow \infty} V_{lj}(r) \equiv 0. 
\eeq
The equivalent but more convenient radial equation takes the form given by 
\beq
  \left[ \frac{d^2}{d\rho^2} - \frac{l(l+1)}{\rho^2} - \frac{V_{lj}(r)}{E} + 1 \right] 
  U_{lj} (\rho) = 0, \label{eq:apC01}
\eeq
where $\rho \equiv kr$ defined with the relative momentum, $k(E) \equiv \sqrt{2E\mu}/\hbar$. 
In numerical calculations, this type of equations can be solved with, 
{\it e.g.} Numerov method explained in Chapter \ref{Ch_3body}. 

To calculate the phase-shift and also other important quantities, 
asymptotic solutions of Eq.(\ref{eq:apC01}) are often necessary. 
In the following, we note these solutions for two major potentials 
frequently used in nuclear physics. 

\subsection{with short-range potential}
Short-range potentials, including nuclear interactions, are characterized as 
\beq
 \lim_{r \rightarrow \infty} V_{lj}(r) < \mathcal{O} (r^{-2}). 
\eeq
The asymptotic condition can be satisfied at $\rho \gg 1$. 
A general solution in this region can be written as 
\beq
 \frac{U_{lj}(\rho)}{\rho} = C_1 j_l(\rho) + C_2 n_l(\rho), 
\eeq
with spherical Bessel and Neumann functions, such as 
\beqa
 j_l (kr) &\longrightarrow & \frac{1}{kr}  \sin \left(kr-l\frac{\pi}{2} \right), \\
 n_l (kr) &\longrightarrow & \frac{-1}{kr} \cos \left(kr-l\frac{\pi}{2} \right). 
\eeqa
Or equivalently, the out-going and in-coming waves can be given as 
\beqa
 h^{(+)}_l (kr) &\equiv & j_l(kr) + i n_l(kr)
    \longrightarrow  \frac{1}{ikr}   e^{ i\left( kr-l\frac{\pi}{2} \right)}, \\
 h^{(-)}_l (kr) &\equiv & j_l(kr) - i n_l(kr)
    \longrightarrow  \frac{-1}{ikr}  e^{-i\left( kr-l\frac{\pi}{2} \right)}. 
\eeqa
Using the coefficients $A_{lj}$ and $B_{lj}$, a general solution takes the form of 
\beqa
 \frac{U_{lj}(kr)}{kr} &=& A_{lj}(E)h^{(+)}_l (kr) + B_{lj}(E)h^{(-)}_l (kr) \nonumber \\
 &=& B_{lj}(E) [S_{lj}(E)h^{(+)}_l (kr) + h^{(-)}_l (kr) ], 
\eeqa
with the S-matrix, $S_{lj}(E) \equiv A_{lj}(E)/B_{lj}(E)$. 
Note that $\abs{S_{lj}(E)}^2 =1$ from the conservation law of the flux. 
Introducing the phase-shift, $\delta_{lj}(E)$ as $S_{lj}(E) \equiv e^{2i\delta_{lj}(E)}$, 
we can get the well-known asymptotic form of $U_{lj}$. 
\beqa
  \frac{U_{lj}(kr)}{kr} &\longrightarrow& \frac{B_{lj}(E)}{ikr} \nonumber 
  \left[ S_{lj}(E) e^{ i\left( kr-l\frac{\pi}{2} \right)} - e^{-i\left( kr-l\frac{\pi}{2} \right)} \right] \\
  && = \frac{B_{lj}(E)e^{i\delta_{lj}(E)}}{ikr} \nonumber 
  \left[ e^{ i\left( kr-l\frac{\pi}{2}+\delta_{lj}(E) \right)} - e^{-i\left( kr-l\frac{\pi}{2}+\delta_{lj}(E) \right)} \right] \\
  && \propto \frac{1}{kr} \sin \left[ kr-l\frac{\pi}{2}+\delta_{lj}(E) \right]. \label{eq:apC05}
\eeqa
Note that $\delta_{lj}(E) \in \mathbb{R}$ since $\abs{S_{lj}(E)}^2 =1$.

\subsection{with Coulomb potential}
It is formulated as 
\beq
 V_{lj}(r) = V(r) = \alpha \hbar c \frac{Z_1 Z_2}{r}, \phantom{00} 
 \alpha \equiv \frac{e^2}{4\pi \epsilon_0 \cdot \hbar c}. 
\eeq
Defining Sommerfeld parameter, $\eta\equiv Z_1 Z_2 \alpha \mu c/\hbar k$, Eq.(\ref{eq:apC01}) 
can be written as 
\beq
  \left[ \frac{d^2}{d\rho^2} - \frac{l(l+1)}{\rho^2} - \frac{2\eta}{\rho} + 1 \right] U_{l} (\rho,\eta) = 0. 
\eeq
With this Coulomb potential, the asymptotic condition can be satisfied at $\rho \gg 2\eta$. 
A general solution takes the form as 
\beq
 \frac{U_{l}(\rho,\eta)}{\rho} = C_1\frac{F_l(\rho,\eta)}{\rho} + C_2\frac{G_l(\rho,\eta)}{\rho}, 
\eeq
where $F_l$ and $G_l$ are the Coulomb functions \cite{72Abramo}. 
Precise derivations of these functions are found in, {\it e.g.} textbook \cite{07Sasakawa}. 
Their asymptotic forms read 
\beqa
 \frac{1}{kr}F_l(kr,\eta) &\longrightarrow & \frac{1}{kr} \sin \left(kr-l\frac{\pi}{2}-\eta\ln 2kr+a_l(\eta) \right), \\
 \frac{1}{kr}G_l(kr,\eta) &\longrightarrow & \frac{1}{kr} \cos \left(kr-l\frac{\pi}{2}-\eta\ln 2kr+a_l(\eta) \right), 
\eeqa
with $a_l(\eta)=\arg \Gamma(l+1+i\eta)$, which is independent of $kr$. 
There is also an iterative formula for $a_l(\eta)$ as 
\beq
 a_{l+1}(\eta) = a_{l}(\eta) + \tan^{-1} \frac{\eta}{l+1}. 
\eeq
Eliminating these unimportant phases, the outgoing and incoming waves 
can be formulated as \cite{07Sasakawa}, 
\beqa
 u^{(+)}_l (kr,\eta) &\equiv & e^{-ia_l(\eta)} \left[G_l(kr,\eta) + i F_l(kr,\eta)\right]
    \longrightarrow e^{ i\left( kr-l\frac{\pi}{2}-\eta\ln 2kr \right)}, \\
 u^{(-)}_l (kr,\eta) &\equiv & e^{ ia_l(\eta)} \left[G_l(kr,\eta) - i F_l(kr,\eta)\right]
    \longrightarrow e^{-i\left( kr-l\frac{\pi}{2}-\eta\ln 2kr \right)}. 
\eeqa
By using these functions, a general solution can be replaced to 
\beqa
 U_{lj}(\rho,\eta) &=& A_{lj}(E,\eta) u^{(+)}_l (kr,\eta) + B_{lj}(E,\eta) u^{(-)}_l (kr,\eta) \\
 &\propto& \left[ S_{lj}(E,\eta) u^{(+)}_l (kr,\eta) + u^{(-)}_l (kr,\eta) \right], 
\eeqa
where we need an additional variable, $\eta$, in two coefficients. 
The S-matrix, $S_{lj}(E,\eta)$, and the phase-shift, $\delta_{lj}(E,\eta)$, can be defined 
similarly in the case with short-range potentials. 
The asymptotic solution is also given as 
\beq
 U_{lj}(\rho,\eta) \longrightarrow \propto 
 \sin \left[ \rho-l\frac{\pi}{2}-\eta \ln 2\rho + \delta_{lj}(E,\eta) \right]. \label{eq:apC06}
\eeq
In the following, however, we will not use Eqs.(\ref{eq:apC05}) and (\ref{eq:apC06}), 
although those are useful for analytic discussions. 

\section{fitting formula for phase shift}
We explain how to compute the S-matrix within the numerical framework. 
First, we consider the position $r=R_b$ at which two particles 
can be separated sufficiently from each other. 
The radial mesh, $dr$, should be enough small compared with $R_b$. 
At this point, we assess the quantity $q$ defined as 
\beq
 q(X) \equiv \frac{U_{lj}(X)}{U_{lj}(X+d)} \label{eq:apC11}
\eeq
with $X\equiv k\cdot R_b$ and $d \equiv k\cdot dr$. 
Remember that the perturbed wave, $U_{lj}(X)$, is computed numerically. 
On the other hand, in the case with Coulomb potential for instance, 
$q(X)$ is also evaluated as 
\beq
 q(X) = \label{eq:apC12}
 \frac{S_{lj}(E,\eta)u^{(+)}_l(X,\eta)+u^{(-)}_l(X,\eta)}{S_{lj}(E,\eta)u^{(+)}_l(X+d,\eta)+u^{(-)}_l(X+d,\eta)}, 
\eeq
where $u_l^{(+)}$ and $u_l^{(-)}$ can be computed independently of $U_{lj}$. 
By solving Eq.(\ref{eq:apC11}) and Eq.(\ref{eq:apC12}) simultaneously for $S_{lj}(E,\eta)$, 
we can get 
\beq
 S_{lj}(E,\eta) = 
 \frac{ U_{lj}(X+d)u_l^{(-)}(X,\eta) - U_{lj}(X)u_l^{(-)}(X+d,\eta) }
      { U_{lj}(X)u_l^{(+)}(X+d,\eta) - U_{lj}(X+d)u_l^{(+)}(X,\eta) }, 
\eeq
and $2i \delta_{lj}(E,\eta) = \ln S_{lj}(E,\eta)$. 
This is the numerical formula for the S-matrix and the phase-shift. 
Notice that the similar formula can be derived in the case with short-range potentials. 
%From the asymptotic formulas, $q_{lj}(E)$ satisfy 
%\beq
% q_{lj}(E) = \frac{\sin[p_{lj}(X,E)+\delta_{lj}(E)] }{\sin[p_{lj}(X-dx,E)+\delta_{lj}(E)] }
%\eeq
%where $p_{lj}(X,E)=X-l\frac{\pi}{2}$ for a short-range potential, 
%whereas $p_{lj}(X,E)=X-l\frac{\pi}{2}-\eta\ln 2X$ for Coulomb potential. 
%Notice that $p_{lj}(X,E)$ can be calculated with unperturbed waves, 
%separately from $U_{lj}(X,E)$. 
%Solving this equation for $\delta_{lj}$ leads to 
%\beq
% \tan\sigma_{lj}(E) = -\frac{\sin p_{lj}(X,E)-q_{lj}(E)\sin p_{lj}(X-dx,E)}{\cos p_{lj}(X,E)-q_{lj}(E)\cos p_{lj}(X-dx,E)}, 
%\eeq

Practically, it is well known that the phase-shift can be fitted by the 
Breit-Wigner distribution. 
That is 
\beq
 \delta_{lj}(E) = \tan^{-1} \left[ \frac{\Gamma_0/2}{E_0-E} \right] + C_{lj}(E), 
\eeq
or equivalently, 
\beq
 \frac{d\delta_{lj}(E)}{dE} = \frac{\Gamma_0/2}{\Gamma_0^2/4 + (E_0-E)^2} + \frac{dC_{lj}(E)}{dE}, \label{eq:apcps}
\eeq
where $C_{lj}(E)$ is a smooth background. 
The central value, $E_0$, and width, $\Gamma_0$, correspond to the complex pole of the S-matrix, 
locating at $E=E_0-i\Gamma_0/2$ \cite{07Sasakawa}. 
%Accordingly, we have got the fitting formula, which is equivalent to Eq.(\ref{eq:sigde}) in Chapter \ref{Ch_Results1}. 

%\include{end}

\renewcommand{\bibname}{References}
%---
%\bibliographystyle{myjpsj3}
%\bibliography{zb_until_2022_1020,zb_until_2023_0831}
%---
%\input{root.bbl}
%---
%b\renewcommand{\bibname}{References-old}
%b\begin{thebibliography}{1}
%===== {Ch_Intro} =====
%b\bibitem{28Har} D.R. Hartree, Proc. Camb. Phil. Soc. {\bf 24}, 89 (1928).
%b\bibitem{30Fock} V.A. Fock, Z. Phys. {\bf 49}, 339 (1930).
%b\bibitem{92Grobe} R. Grobe and J.H. Eberly, Phys. Rev. Lett. {\bf 68}, 2905 (1992). %soft Coulomb, auto-ionization.
%---

\end{document}